\documentclass[reprint,amsmath,amssymb,aps,prb]{revtex4-2}
\allowdisplaybreaks

\usepackage{hyperref}
\usepackage{lineno}
\usepackage{graphicx}
\usepackage{dcolumn}
\usepackage{bm}
\usepackage{xcolor}
\usepackage[utf8]{inputenc}
\usepackage{siunitx}

\usepackage{lmodern}
\usepackage{import}
\usepackage{layouts}

\usepackage{acronym}
\usepackage{float}
\usepackage{booktabs}
\usepackage{array} 
\usepackage{calc} 
\usepackage{rotating}

\renewcommand{\vec}[1]{\textit{\textbf{#1}}}

\newcommand{\dxx}{\,\text{d}\vec{x}}

\newcommand{\diff}[2]{\frac{\text{d}#1}{\text{d}#2}}

\newcommand{\pdiff}[2]{\frac{\partial #1}{\partial #2}}

\newcommand{\vdiff}[2]{\frac{\delta #1}{\delta #2}}

\newcommand{\vecs}[2]{\vec{#1}_{\text{#2}}}
\newcommand{\Energy}[1]{E_{\text{#1}}}

\newcommand{\ssub}[2]{#1_{\text{#2}}}
\newcommand{\brak}[1]{\left( #1 \right)}
\newcommand{\bbrak}[1]{\big( #1 \big)}
\newcommand{\Brak}[1]{\left[ #1 \right]}

\newcommand{\Ediff}[1]{\diff{\Energy{#1}}{t}}

\renewcommand{\exp}[1]{\text{exp}\!\brak{#1}}
\renewcommand{\ln}[1]{\text{ln}\!\brak{#1}}
\renewcommand{\log}[1]{\text{log}_{10}\!\brak{#1}}
\renewcommand{\sin}[1]{\text{sin}\!\brak{#1}}
\renewcommand{\cos}[1]{\text{cos}\!\brak{#1}}

\renewcommand{\max}[1]{\text{max}\!\brak{#1}}

\newcommand{\sinkx}{\sin{k x - \omega t}}
\newcommand{\coskx}{\cos{k x - \omega t}}

\newcommand{\Hext}{\ssub{H}{ext}}
\newcommand{\phiH}{\ssub{\phi}{H}}
\newcommand{\uxt}{\vec{u}(\vec{x}, t)}
\newcommand{\dSij}{\Delta S_{ij}}
\newcommand{\dSTO}{\Delta \ssub{S}{21}}

\newcommand{\Ms}{\ssub{M}{s}}

\newcommand{\Ku}{\ssub{K}{u}}

\newcommand{\IntO}[1]{\int_{\Omega} #1 \dxx}

\newcommand{\kl}{\ssub{\kappa}{l}}
\newcommand{\kt}{\ssub{\kappa}{t}}
\newcommand{\vl}{\ssub{v}{l}}
\newcommand{\vt}{\ssub{v}{t}}

\begin{document}

\title{
    Magnetically Programmable Surface Acoustic Wave Filters: \texorpdfstring{\\}{ }
    Device Concept and Predictive Modeling
}

\author{Michael K. Steinbauer$^{1,2,3,*}$}
\author{Peter Flauger$^{1,2}$}
\author{Matthias Küß$^4$}
\author{Stephan Glamsch$^4$}
\author{Emeline D. S. Nysten$^5$}
\author{Matthias Weiß$^5$}
\author{Dieter Suess$^{1,2}$}
\author{Hubert J. Krenner$^5$}
\author{Manfred Albrecht$^4$}
\author{Claas Abert$^{1,2}$}

\affiliation{
$^1$University of Vienna, Faculty of Physics, Physics of Functional Materials, 1090 Vienna, Austria\\
$^2$University of Vienna, Research Platform MMM Mathematics-Magnetism-Materials, 1090 Vienna, Austria\\
$^3$University of Vienna, Vienna Doctoral School in Physics, 1090 Vienna, Austria\\
$^4$University of Augsburg, Institute of Physics, 86135 Augsburg,
Germany \\
$^5$Universität Münster, Physikalisches Institut, 48149 Münster, Germany \\
*Corresponding author; e-mail address: michael.karl.steinbauer@univie.ac.at
}

\date{\today}

\begin{abstract}

\begin{center}
\bfseries \large Abstract
\end{center}

Filtering surface acoustic wave (SAW) signals of specified frequencies depending on the strength of an external magnetic field in a magnetostrictive material has garnered significant interest due to its potential scientific and industrial applications. Here, we propose a device that achieves selective SAW attenuation by instead programming its internal magnetic state. To this end, we perform micromagnetic simulations for the magnetoelastic interaction of the Rayleigh SAW mode with spin waves (SWs) in exchange-decoupled Co/Ni islets on a piezoelectric LiTaO$_3$ substrate. Due to the islets exhibiting perpendicular magnetic anisotropy, the stray-field interaction between them leads to a shift in the SW dispersion depending on the magnetic alignment of neighboring islets. This significantly changes the efficiency of the magnetoelastic interaction at specified frequencies. We predict changes in SAW transmission of \SI{52.0}{dB/mm} at \SI{3.8}{GHz} depending on the state of the device. For the efficient simulation of the device, we extend a prior energy conservation argument based on analytical solutions of the SW to finite-difference numerical calculations, enabling the modeling of arbitrary magnetization patterns like the proposed islet-based design.

\end{abstract}
                         
\maketitle

\section*{\label{sec:Introduction} Introduction}

\begin{figure*} [!htbp]
    \centering
        \includegraphics[width=0.97\textwidth]{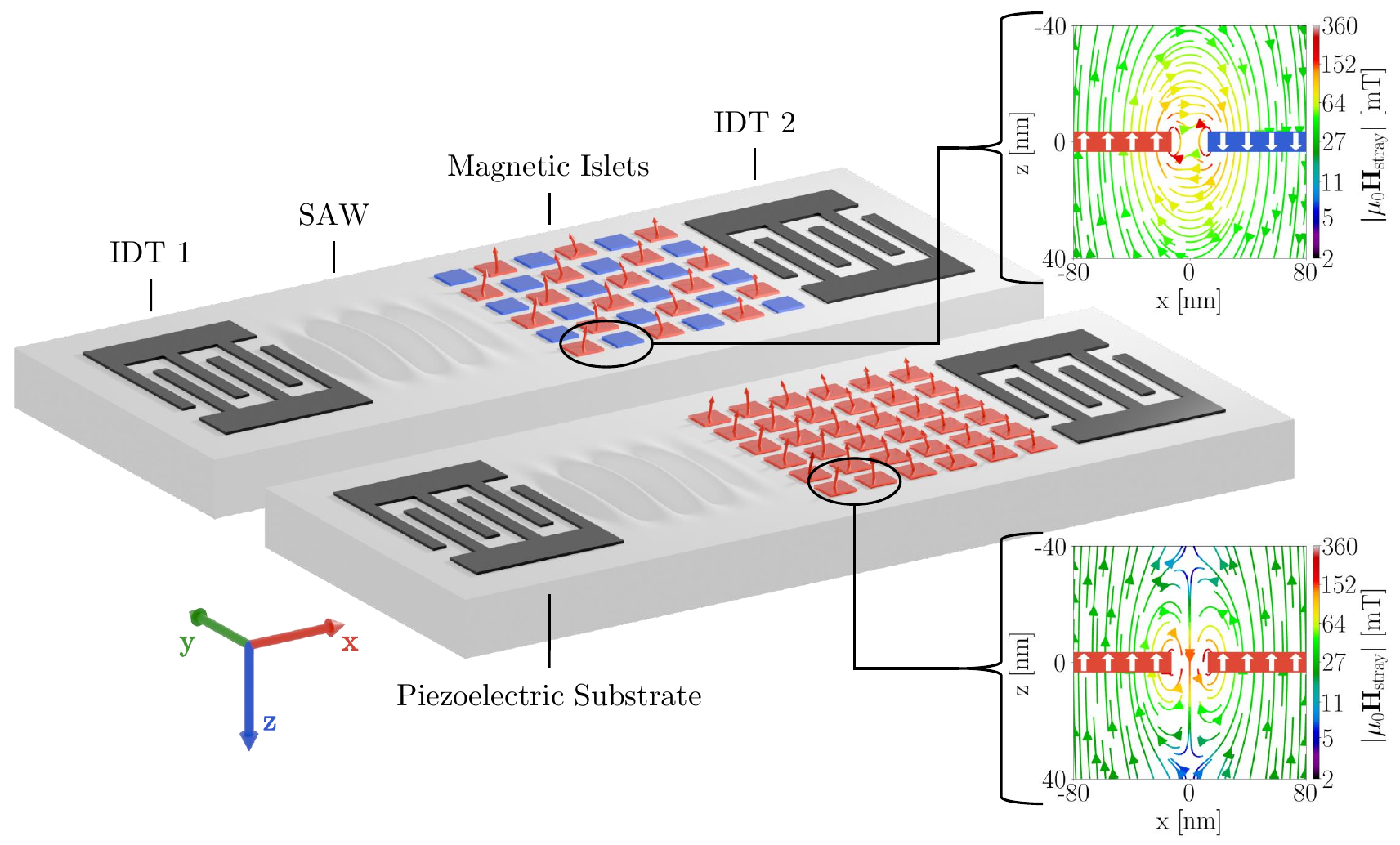}
        \caption{Illustration of the proposed device in antiparallel configuration (top) and parallel configuration (bottom). Insets show cross-sections of the stray-fields between neighboring islets. Some simplifications were made in this depiction to show the principal functionality of the device more clearly: a reduction in the number of islets and increase to their size, exaggeration of SAW amplitude, the omission of the capping layer and constant external bias field, as well as assuming a homogeneous magnetization of the islets.}
        \label{fig:DeviceConcept}
\end{figure*}

Surface acoustic wave (SAW) devices are integral to the modern telecommunications infrastructure, where they have already functioned as high-performance band-pass filters for decades \cite{Morgan1998SAW, Morgan2010SAW, Campbell2012SAW}. The foundational principle for their application as filters was established with the introduction of specially shaped electrodes, called interdigital transducers (IDTs), on piezoelectric substrates \cite{White1965IDTOriginal}.

An alternating voltage applied to a transmitting IDT excites SAWs within the substrate. These waves propagate across the substrate to a receiving IDT, where they are converted back to an electrical signal. The geometric configuration of the IDT dictates the resonance frequency at which SAWs can be excited, along with higher-order harmonics. This inherent frequency selectivity enables the precise bandpass filtering characteristic of SAW devices, while their small size and operating range in the radio frequency (RF) regime make them ideal for use in hand-held devices like smartphones \cite{Ruppel2002SAW, Chen2022SAWFilter, Liu2025SAWFilters5G}.

While a different kind of filter, the bulk acoustic wave (BAW) filter, is currently dominant for high-frequency applications in the telecommunications industry \cite{Liu2020BAW, Yang2022Trends}, SAW devices remain relevant.
This is because the basic functionality of SAW devices can be extended by modifying the substrate between the IDTs in various ways. This modification enables their application as lab-on-a-chip platforms for advancements across diverse disciplines, including physics, chemistry, and medicine \cite{Delsing2019SAWRoadmap, Mandal2022SAWRoadMap, Ding2012Microfluidics, Rufo2022Biomedical}. A significant area of interest involves the interaction of these devices with magnetism. Magneto-elastic coupling is historically well-established for its ability to alter the properties of a SAW \cite{Webb1979, Yamaguchi1980}. Recently, significant focus has been put on these interactions in the RF regime \cite{luo2024MagnetoElectricRoadMap, Bozhko2020MagnonReview}. Specifically, the excitation of spin waves (SWs) via the magnetoelastic effect has emerged as a promising tool in magnonics, offering numerous potential applications \cite{Yang2021MagnetoElectricReview, Puebla2022MagnetoElectricReview}. One such application is the ability to absorb specific frequencies, dependent on the strength of an applied external magnetic field, when the SAW interacts with a magnetic thin film \cite{Weiler2011FMRTheory, Dreher2012Theory, Li2017Spinwave, Casals2020SWTransportSAW, Kunz2024Spinwave}. 

This can be achieved, because SAW attenuation is only significant at frequencies where its dispersion relation intersects with that of the SW. There, the SAW's energy is transduced to a SW, where it then dissipates via Gilbert damping \cite{Yamamoto2022DRMatch, Kuess2021Experiment, Bas2022DRMatch}. The crossover point of the dispersion relations can be precisely tuned by adjusting the strength of an external magnetic field, as the field shifts the SW dispersion relation, thereby repositioning the resonance point to a desired frequency. By tuning this point to the band-pass frequency, signals are attenuated, whereas when detuned, the signal passes without attenuation, resulting in highly efficient signal filtering. However, implementing such a device would necessitate the continuous operation of a variable electromagnet, which is undesirable for many potential applications.

In this study, we propose a different method to deliberately shift the SW dispersion, where we utilize exchange-decoupled, but stray-field interacting, thin magnetic islets with out-of-plane (OOP) magnetization and a large magnetic moment (see Fig. \ref{fig:DeviceConcept}). The magnetization direction of each islet, influenced by the device's magnetic history, can align along either the positive or negative z-axis. This orientation significantly alters the SW dispersion relation, which in turn impacts SAW absorption. This inherent variability allows for diverse device configurations to manipulate both the amount of absorption and the frequency band of its occurrence. Once programmed via a variable external magnetic field or spin torque methods \cite{Guo2021STTSOTRoadmap}, the islets maintain a stable magnetic state, requiring only a constant bias field to facilitate magnetoelastic interaction. This would reduce both size and energy requirements significantly compared to existing field-tuned solutions, while offering similar capabilities.

We compare two device configurations: the A-state, where adjacent islets have antiparallel magnetic moments, and the P-state, where they are parallel. In the A-state, stray fields form flux closures, which enhance the internal magnetic fields within the islets. This increases the system's stiffness and, consequently, its resonance frequencies compared to islets in the P-state, where magnetic field lines repel. Therefore, we anticipate a shift in the SW dispersion relation between the P- and A-states, leading to differences in SAW absorption, which we test using the Rayleigh SAW mode as an example.

\section*{Results}

\subsection*{Micromagnetics}

In order to explore the feasibility of the proposed device, we make use of micromagnetism, which enables the simulation of spatially varying magnetization patterns, a crucial requirement due to the islet-based architecture of the device. The equation of motion of the magnetization vector is given by the Landau-Lifshitz-Gilbert equation (LLG) \cite{LL1935LLG, Gilbert1955LLG}

\begin{align} \label{eq:LLG}
    \diff{\vec{m}}{t} = \frac{-\gamma}{1+\alpha^2}\Brak{\vec{m}\times\vecs{H}{Eff} + \alpha\vec{m}\times\brak{\vec{m}\times\vecs{H}{Eff}}},
\end{align}

\noindent where $\gamma = \mu_0 \gamma_e$ is the reduced gyromagnetic ratio with the vacuum permeability $\mu_0$, $\vec{m} = \vec{M}/\Ms$ is the reduced magnetization with the saturation magnetization $\Ms$ and $\alpha$ is the Gilbert damping. The effective field $\ssub{\vec{H}}{eff}$ is given by the negative variational derivative of the total magnetic energy $\ssub{E}{M}$ with respect to the magnetic polarization. $\ssub{E}{M}$ can be comprised of several energy contributions in the magnetic region(s) $\Omega$, such as the Zeeman energy under an external field $\vecs{H}{ext}$, stray-field energy with demagnetization field $\ssub{\vec{H}}{dem}$, exchange energy with exchange stiffness $\ssub{A}{ex}$, and uniaxial magnetic anisotropy with strength $\Ku$ along the easy axis $\ssub{\vec{e}}{u}$ \cite{Abert2019MicroMagReview}. To describe the interaction of the SAW with the magnetic islets of the device, $\ssub{E}{M}$ also has to include a magneto-elastic energy term $\ssub{E}{MagEl}$ describing the coupling of the normalized magnetization $\vec{m}$ with the displacement of the material $\vec{u}$. This term is given by \cite{Shu2004MagElMicroMag, Zhang2005MagnetoElasticEnergy, Chikazumi2009PolyChryst, Liang2014Theory}

\begin{align}
    \ssub{E}{MagEl} &= \frac{1}{2} \IntO{\brak{\ssub{\boldsymbol{\varepsilon} - \boldsymbol{\varepsilon}}{m}}  : C : \brak{\ssub{\boldsymbol{\varepsilon} - \boldsymbol{\varepsilon}}{m}}}\label{eq:DefinitionMagElEnergy} \\ 
    \varepsilon_{ij} &= \frac{1}{2} \brak{\pdiff{u_i}{x_j} + \pdiff{u_j}{x_i}} \label{eq:DefinitionRegularStrain} \\
    \varepsilon_{\text{m}, ij} &=
    \begin{cases}
        \frac{3}{2} \ssub{\lambda}{s}(m_i^{2}-\frac{1}{3})\quad i = j\\
        \\
        \frac{3}{2} \ssub{\lambda}{s}m_i m_j \quad \quad \, \, \, i \neq j \label{eq:DefinitionMagneticStrain}
    \end{cases}
\end{align}

\noindent where $C$ is the stiffness tensor, $\ssub{\lambda}{s}$ the saturation magnetostriction, $\boldsymbol{\varepsilon}$ the strain associated with the displacement, $\ssub{\boldsymbol{\varepsilon}}{m}$ the strain associated with the local magnetization and ":" denotes the Frobenius inner product. In Eq. (\ref{eq:DefinitionMagneticStrain}), a polycrystalline material with at least cubic symmetry was assumed. $\ssub{E}{M}$ and $\ssub{\vec{H}}{eff}$ can then be expressed by

\begin{align}
    \ssub{E}{M} =
    &\int_{\Omega}-\mu_0 M_s \vec{m}\cdot \Brak{\vecs{H}{Ext} + \frac{1}{2}\vecs{H}{dem}} \notag \\
    &+ \ssub{A}{ex} \sum_{i, j} \brak{\pdiff{m_i}{x_j}}^2 
    - \Ku \brak{\vec{m} \cdot \ssub{\vec{e}}{u}}^2 \notag \\
    &+\frac{1}{2} \brak{\ssub{\boldsymbol{\varepsilon} - \boldsymbol{\varepsilon}}{m}}  : C : \brak{\ssub{\boldsymbol{\varepsilon} - \boldsymbol{\varepsilon}}{m}} \dxx \label{eq:DefEnergy} \\
    \ssub{\vec{H}}{eff} = &- \frac{1}{\mu_0 \Ms} \vdiff{\ssub{E}{M}}{\vec{m}}. \label{eq:Heff}
\end{align}

Self-consistent solvers, capable of simulating the coupled magnetic and mechanical dynamics from initial- and boundary conditions of the magnetization, displacement, and momentum can be found in ref. \cite{Liang2014Theory} for finite-element models and ref. \cite{Vanderveken2021FullyCoupMuMax, Flauger2025Modeling} for finite-difference models. However, these self-consistent approaches are computationally intensive. This is because they need to solve multiple coupled differential equations each timestep. Crucially, they must also simulate large parts or even the entire substrate. Because the substrate is often significantly thicker than the magnetic region, this can increase the system size by multiple orders of magnitude, leading to substantially increased simulation times.

\subsection*{Uni-Directional Model}

To mitigate this problem and efficiently calculate the magnetic transmission losses $\dSij$ of a SAW signal due to interaction with the magnetic islets of the device, we augment the macrospin model developed in ref. \cite{Weiler2011FMRTheory, Dreher2012Theory, Kuess2020DRFormula} with micromagnetic simulations performed with the finite-difference python library magnum.np \cite{Bruckner2023MagnumNP}. This hybrid model was validated on experimental results from ref. \cite{Kuess2021Experiment}. The validation as well as the full derivation of the model can be found in the Methods section. A short overview of the working principle is given here: 

The losses $\dSij$ in decibels are given by

\begin{align}
    \dSij = 10\text{log}_{10}\brak{\frac{\ssub{P}{out}}{\ssub{P}{in}}},
\end{align}

\noindent where $\ssub{P}{in}$ and $\ssub{P}{out}$ are the SAW power before and after interaction with the device, respectively. Note that $\dSij$ only takes into account the magnetoelastic interaction and disregards other contributions to the total transmission losses $S_{ij}$. In the model, the displacement $\vec{u}$ drives magnetization processes as part of the LLG according to Eq. \ref{eq:LLG}-\ref{eq:Heff}. However, the energy the SAW loses to the magnetic system during this process is assumed to result solely in a decay of its amplitude, which gets estimated from an energy conservation argument. The magnetic system gains energy in this way until an equilibrium is reached, where the energy pumped into the system by the SAW equals the energy lost due to Gilbert damping.

Unlike macrospin-based magneto-elastic models \cite{Weiler2011FMRTheory, Dreher2012Theory, Kuess2020DRFormula}, this approach is capable of predicting the results even for complex magnetization textures like the islet design, as no analytical solutions for the spin wave dynamics have to be provided. Namely, given the parameterization of some SAW mode with velocity $c$, we simulate the rate of energy transfer from the elastic into the magnetic system, $\ssub{R}{T}$, over a few periods of the SAW until this energy flow becomes constant ($\diff{}{t} \ssub{R}{T} = 0$). From this value and the total SAW energy $\ssub{E}{Ph}$, obtained in a pre-simulation step, we can estimate the spatial rate for the energy loss of the SAW, allowing us to determine $\dSij$ for a signal that has traversed the islet pattern of length $l$: 

\begin{align}
    \dSij(l) &= \frac{10}{\ln{10}} \frac{l}{c} \frac{-\ssub{R}{T}}{\Energy{Ph}} \label{eq:dSijCalcTheory} \\
    \ssub{R}{T} &=  -\IntO{\brak{\pdiff{}{t} \boldsymbol{\varepsilon}} : C : \boldsymbol{\varepsilon}_{\text{m}}} \label{eq:RTTheory}\\
    \Energy{Ph} &= \frac{1}{2} \int_{\text{V}} \boldsymbol{\varepsilon} : C : \boldsymbol{\varepsilon} + \rho \vec{v}^{2} \dxx. \label{eq:SAWEnergy}
\end{align}

\noindent Here, $V$ is the volume of the simulated section, $\Omega$ are the magnetic region(s) of that volume, $\rho$ is the material density, and $\vec{v} = \diff{\vec{u}}{t}$. Because only $\ssub{R}{T}$ has to be determined in simulation, but vanishes outside $\Omega$, this reduces the computational complexity from 3D to quasi-2D for thin magnetic structures, like is the case with the islet design, leading to a significant increase in computational speed.

 \subsection*{\label{sec: Material Parameters}Material Parameters and Device Geometry}
 
 Co/Ni multilayer structures were identified as a possible material candidate for the magnetic islets of the programmable device. They have shown OOP uniaxial magnetic anisotropy, a sizable magnetic moment \cite{You2012CoNiMultiLayer, He2021CoNiGilbert, Mizukami2011CoNiGilbert}, which is necessary in order to maximize the stray-field interaction between the islets themselves, and a moderate Gilbert damping \cite{He2021CoNiGilbert, Mizukami2011CoNiGilbert}. The parameters used in this section are based on a Ta(\SI{3}{nm})/Pt(\SI{20}{nm})/[Co(\SI{0.4}{nm}/Ni(\SI{0.3}{nm})]$_{10}$/ Ta(\SI{3}{nm})/Pt(\SI{3}{nm}) film studied in ref. \cite{You2012CoNiMultiLayer}. While the magnetostriction coefficient $\ssub{\lambda}{s}$, necessary to facilitate efficient phonon-magnon coupling, was not measured in ref. \cite{You2012CoNiMultiLayer}, similar multilayer structures have shown substantial $\ssub{\lambda}{s}$ values \cite{Gopman2018MagnetostrictionCoNiMultilayers}. $\ssub{\lambda}{s}$, as well as other properties not directly available in the literature, were estimated from the individual bulk values of the materials. The parameters used for the simulation of the magnetic Co/Ni multilayer are given in Table \ref{tab:DeviceParameters} \cite{He2021CoNiGilbert, Mizukami2011CoNiGilbert, Klokholm1982NiLambLit, Kim2018CoLambLit, You2012CoNiMultiLayer, MaterialCobalt, MaterialNickel, Jain2013MaterialsProject}. To prevent the islets from coupling via exchange interaction, we separate them with a demagnetized spacer material where $\ssub{A}{ex} = \vec{m} = \Ku = 0$. To simplify the model, we assume the mechanical properties of both spacings and islets to be identical and isotropic. Depending on how the spacings are realized experimentally (see sec. Experimental Feasibility), they could influence the SAW mode however.

\begin{table}[H] 
    \centering 
    \caption{Assumed material parameters used for the simulation of the magnetic Co/Ni multilayer, based on a Ta(\SI{3}{nm})/Pt(\SI{20}{nm})/[Co(\SI{0.4}{nm}/Ni(\SI{0.3}{nm})]$_{10}$/ Ta(\SI{3}{nm})/Pt(\SI{3}{nm}) film. \cite{He2021CoNiGilbert, Mizukami2011CoNiGilbert, Klokholm1982NiLambLit, Kim2018CoLambLit, You2012CoNiMultiLayer, Jain2013MaterialsProject, MaterialCobalt, MaterialNickel}} 
    \begin{tabular}{lcccccccc} 
        \toprule 
        
        $\ssub{A}{ex}$ & $\ssub{M}{s}$ & $\Ku$ & $\alpha$ & $\ssub{\lambda}{s}$ & $\rho$ & $E$ & $\nu$ \\
        
        $[\text{pJ/m}]$ & $[\text{kA/m}]$ & $[\text{kJ/m}^3]$ & $[1]$ & $[10^{-6}]$ & $[\text{kg/m}^3]$ & $[\text{GPa}]$ & $[1]$ \\
        
        \midrule 10 & 725 & 366 & 0.05 & -51.71 & 9141 & 270.7 & 0.263 \\ 
        
        \bottomrule  
        
    \end{tabular} \label{tab:DeviceParameters}
\end{table}

\noindent Here, $E$ is the Young modulus and $\nu$ the Poisson ratio. The uniaxial magnetic anisotropy $\Ku$ along the OOP direction is larger than the thin film shape anisotropy ($\frac{1}{2} \mu_0 \Ms^2$) giving rise to an effective OOP easy axis direction.

\begin{figure}[H]
    \centering
        \includegraphics[width=0.47\textwidth]{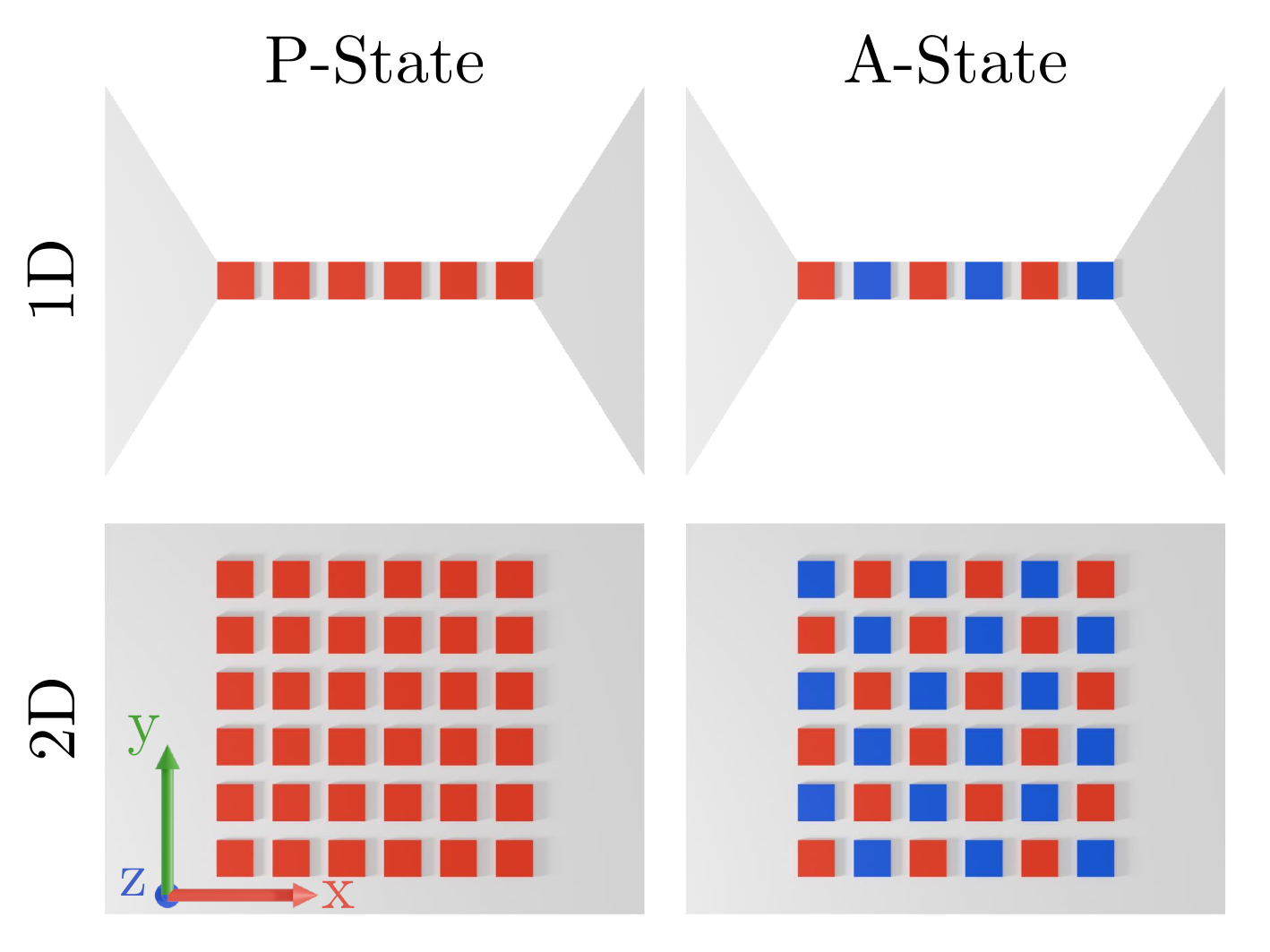}
    \caption{Illustration of the studied islets in parallel (P) and antiparallel (A) configurations in a 1D and 2D device: Red signifies a magnetization pointing OOP away from the substrate, blue into the substrate.}
    \label{fig:IsletConfigs}
\end{figure}

Each islet has a size of $200 \times 200 \times 36\,\si{nm^3}$ with a spacing of \SI{25}{nm} between them and is placed on a 36°-rotated Y-cut X-propagation LiTaO$_3$ substrate. We investigate two distinct configurations of these islets (see Fig. \ref{fig:IsletConfigs}): A 1D device designed to characterize the behavior of islets situated atop a waveguide and a 2D device representative of a conventional SAW filter. Due to the increased number of neighbors in the 2D device, we expect the shift in dispersion relation to be more pronounced compared to the 1D case. Possible avenues for the experimental realization of the device are discussed in the Methods section.

To model these configurations, islets are arranged in a line and periodic boundary conditions (PBCs) in x direction are applied for the 1D device, whereas for the 2D device, they are arranged in an $n \times 2$ grid with PBCs in both the x and y directions. The application of PBCs is essential for accurate device modeling, as it mitigates finite-size effects of the stray field, thereby enabling the simulation of the whole device from a much smaller sub-region. For both devices, the discretization of the simulation geometry is (\SI{5}{nm}, \SI{5}{nm}, $\frac{7}{4}\,\si{nm}$) when simulating magnetization dynamics, where only the \SI{7}{nm} thick Co/Ni multilayer is considered.

 \subsection*{\label{sec: Dispersion Relations} Dispersion Relation}

In order to quickly evaluate different sets of material parameters and geometries in regards to their programmability, we calculate the spin wave dispersion for the parallel and antiparallel state. We employ the methodology proposed in ref. \cite{Venkat2013Disp}, where a magnetic sinc-pulse $\vecs{h}{excite}$ in space and time is used to excite all possible SW modes at once. In addition, the OOP uniaxial magnetic anisotropy $\Ku$, a constant bias field $\mu_0 \ssub{H}{ext} = 50\,\si{mT}$ applied along the x-axis and the demagnetization and exchange energies are considered. The external field will later be necessary to facilitate efficient magneto-phononic coupling, because the main strain components of the Rayleigh SAW mode, $\ssub{\varepsilon}{xx}$ and $\ssub{\varepsilon}{zz}$, favor an equilibrium magnetization which is tilted towards the x-axis \cite{Kuess2022}. Additionally, the field is used to shift the spin wave dispersion into a desired frequency range, with higher fields leading to lower resonance frequencies. It could be provided by a bias magnet in an experimental realization.

In this simulation, the 1D and 2D magnetic islet patterns are 24 islets long (\SI{5.4}{\micro\meter}). First, they are relaxed using a Gilbert damping $\alpha$ of 1 and without $\vecs{h}{excite}$. Due to the demagnetization energy and the external bias field, the equilibrium magnetization of the islets is not homogeneous. While the magnetization at the edges of the islets remains mostly parallel to the z-axis, the x-component of the magnetization increases towards the islet center with the polar angle depending on the device and state. For the parameters described in sec. Material Parameters and Device Geometry, the polar angle at the islet center is $\approx$20.0° and $\approx$21.5° for the A- and P-states of the 1D device respectively, and $\approx$19.5° and $\approx$24.0° for the A- and P-states of the 2D device.

\begin{figure}
    \centering
        \includegraphics[width=0.47\textwidth]{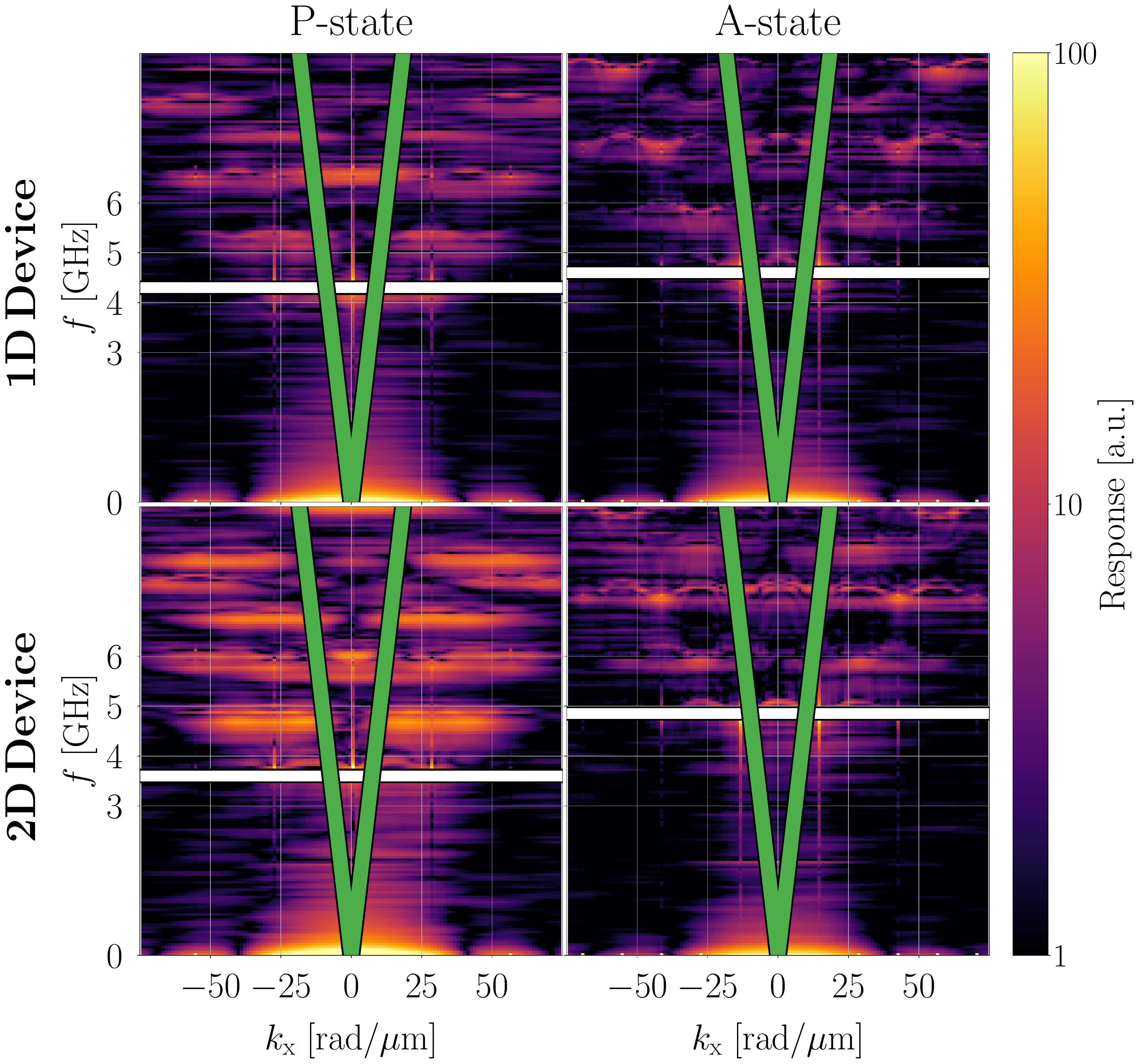}
    \caption{Magnetization dynamics response of the 1D and 2D devices described in sec. Material Parameters and Device Geometry in the P- and A-states to a magnetic sinc pulse under the influence of a \SI{50}{mT} bias field along the x direction. The scales of the plot were limited to the range of 1 to 100 to emphasize the relevant data. The linear SAW dispersion relation for $c = 3068.2\,\si{m/s}$ is overlaid as a green line. The frequencies where the SAW dispersion intersects the first order SW dispersion are highlighted by white lines: \SI{4.30}{GHz} and \SI{4.60}{GHz} for the P- and A-states, respectively, of the 1D device; \SI{3.60}{GHz} and \SI{4.85}{GHz} for the 2D device.}
    \label{fig:Dispersion}
\end{figure}

The simulation is then performed for \SI{20}{ns} with \SI{1}{ps} integration steps and $\alpha = 10^{-8}$. The Gilbert damping was set close to zero in order to sharpen the dispersion relation. During this, the magnetization gets saved for each time step, from which the dispersion relation gets calculated via a Fast Fourier Transform. The simulation results for the devices described in sec. Material Parameters and Device Geometry are presented in Fig. \ref{fig:Dispersion}: The islet pattern magnonic crystal \cite{Krawczyk2014MagnonCrystalDispersion} shows a complicated discretized dispersion relation which largely differs for the P- and A-state. As expected, the shift in the SW dispersion for the 2D device (\SI{1.25}{GHz}) is larger than that of the 1D device (\SI{0.3}{GHz}). We attribute the quite broad resonances to both stray fields and equilibrium magnetization not being homogeneous across the islet.

\subsection*{\label{sec: SAW parameterization}SAW Parameterization}

While LiTaO$_3$ can support both shear horizontal and Rayleigh SAW modes \cite{Nakamura1977LiTaO3}, this study will focus on the Rayleigh mode. Because this mode exhibits non-vanishing displacement both in the direction of travel and transversal to it \cite{Sonner2021RayleighTransversalComponent}, it has a sense of rotation, leading to a non-reciprocal absorption of the SAW depending on the direction of travel under an external field \cite{Dreher2012Theory, Kuess2021Experiment}. This non-reciprocity will be used to validate the uni-directional model in the Methods section.

Using the convention of a z-axis which starts at the surface and points into the device, a parameterization of the Rayleigh mode is given in ref. \cite{Maekawa1976SAWParamet} as: 

 \begin{align}
     \vec{u} = \, &A \frac{\Tilde{\vec{u}}}{\max{|\Tilde{\vec{u}}|}} \label{eq:NormalizeRay} \\
     \ssub{\Tilde{u}}{x}(x, z, t) = \, &\kt \coskx \notag \\
     &\cdot \Brak{\exp{-\kt z} - \frac{2k^2}{k^2+\kt^2} \exp{-\kl z}} \label{eq:uxMaekawa} \\
     \ssub{\Tilde{u}}{y}(x, z, t) = \, &0 \\
     \ssub{\Tilde{u}}{z}(x, z, t) = \, &-k \sinkx \notag \\
     &\cdot \Brak{\exp{-\kt z} - \frac{2 \kt \kl}{k^2+\kt^2} \exp{-\kl z}} \label{eq:uzMaekawa}
 \end{align}

\noindent where $A$ is the amplitude of the SAW, with $\kl$ and $\kt$ being the longitudinal- and transversal penetration components. They depend on the angular frequency $\omega$ and wave number $k$ of the SAW, as well as the longitudinal- and transversal velocities $\vl$ and $\vt$:

  \begin{align}
    \label{eq:DefVL}
    \kl &= \sqrt{k^2 - \frac{\omega^2}{\vl^2}}\\
    \label{eq:DefVT}
    \kt &= \sqrt{k^2 - \frac{\omega^2}{\vt^2}}.
 \end{align}

These equations were derived for a single isotropic material. However, we assume they still model the full device well, because the islets' thickness of \SI{36}{nm} is significantly lower than the employed SAW wavelengths (\SI{511}{\nano \meter} to \SI{1023}{\nano \meter}), resulting in the majority of the SAW energy being confined to the substrate rather than the islets. The dispersion relation is then given by \cite{Landau1960TOE}

 \begin{align} \label{eq:DispersionRelationSAW}
     \omega = \vt \xi k,
 \end{align}

\noindent where we identify $\vt \xi$ to be the velocity $c$ of the SAW with $\xi$ being the positive, real valued solution of the governing equation \cite{Landau1960TOE}

 \begin{align} \label{eq:Xi}
     \xi^6 - 8 \xi^4 + 8\xi^2\brak{3 - 2 \frac{\vt^2}{\vl^2}} - 16 \brak{1 - \frac{\vt^2}{\vl^2}} = 0.
 \end{align}

 To determine $\vt$ and $\vl$, a finite-element simulation of the layered system was carried out using COMSOL\textsuperscript{\textregistered} \cite{COMSOL} at $f=3.07\,\si{GHz}$ where the resulting depth profile of $\uxt$ was fitted to Eq. \ref{eq:uxMaekawa} and \ref{eq:uzMaekawa}. $\uxt$ is thus parametrized by the values in Table \ref{tab:SAWParams} for Eq. \ref{eq:NormalizeRay} to \ref{eq:Xi}. Small frequency dependent changes in $c$ were neglected and $A_0$ was chosen such that it is large enough to minimize numerical errors, while also being small enough to stay in the linear regime of the LLG. For material parameters similar to the ones employed here, the validity range of $A_0$ is approximately $10^{-15}$ m to $10^{-10}$ m. Furthermore, the modeling assumes weak coupling, as observed in experiments for similar continuous magnetic films \cite{Dreher2012Theory, Kuess2021Experiment}.
 
\begin{table}[H]
\centering
    \caption{SAW parameters}
    \begin{tabular}{lccc}
        \toprule
        $c$ & $v_l$ & $v_t$ & $A_0$ \\
        $[\text{m/s}]$ & $[\text{m/s}]$ & $[\text{m/s}]$ & $[\text{pm}]$ \\
        \midrule
        3068.2 &  5312.4 &  3398.3 & 4.00 \\
        \bottomrule
    \end{tabular}
    \label{tab:SAWParams}
\end{table}

\subsection*{\label{sec: Micromagnetic Simulations} Magneto-Phononic Interaction}

To calculate the SAW transmission of our devices, we sequentially excite them with SAWs between \SI{3}{GHz} and \SI{6}{GHz}. The frequencies used, $\ssub{f}{SAW}$, were chosen such that $n$ of their corresponding wavelengths (ranging from \SI{511}{\nano \meter} to \SI{1023}{\nano \meter}) are equal to $2m$ islets and spacings, where $n$ and $m$ are integers:

\begin{align}
     n \frac{c}{\ssub{f}{SAW}} \label{eq:SAWCondition} =  2m \cdot (200\,\si{nm} + 25\,\si{nm}).
\end{align}

\noindent This is necessary to ensure that PBCs can be used. The simulated section of the magnetic pattern then consists of $2m$ islets and buffers. Note that this section can be much smaller than the total length of the islet pattern ($\approx$\SI{1}{mm}). For each design, configuration, and frequency, the following procedure was then repeated: The initial magnetization configuration of the islets was set along the OOP axis and neighboring islets are initialized either with parallel or antiparallel magnetization, depending on the state of the device, after which this state was relaxed at $\alpha = 1$ under the influence of (i) demagnetization energy, (ii) exchange energy with stiffness $\ssub{A}{ex}$, (iii) OOP uniaxial anisotropy with strength $\Ku$ and (iv) Zeeman energy of a \SI{50}{mT} external field along the x-axis. For the simulations themselves, $\alpha = 0.05$ was set and the (v) magneto-elastic energy with saturation magnetization $\ssub{\lambda}{s}$ was added to energy terms (i) to (iv). Then, the SAW was activated and the simulation proceeded with a time discretization of $\Delta t = \frac{1}{50 \ssub{f}{SAW}}$. At each time step, $\dSTO$ was calculated according to Eq. \ref{eq:dSijCalcTheory} for a \SI{1}{mm} long islet pattern where $\Energy{Ph}$ was determined in a pre-simulation step using  COMSOL\textsuperscript{\textregistered} \cite{COMSOL} and scaled according to $\Energy{Ph} \propto f$. The simulation concludes when, for 50 consecutive time steps (one period of the SAW), the relative change in $\dSTO$ is less than $10^{-5}$ or its absolute change is less than $10^{-3}\,\si{dB}$:

\begin{align} 
    \label{eq:ValidationAbortRel} 
    \left|  \frac{\dSTO(t) - \dSTO(t - \Delta t)}{\dSTO(t)} \right| &< 10^{-5}\\
    \label{eq:ValidationAbortAbs}
    \left| \, \dSTO(t) - \dSTO(t - \Delta t) \, \right| &< 10^{-3}\,\si{dB}.
\end{align}

\noindent This convergence criterion, while equivalent to requiring a constant energy flow because $\dSTO \propto \ssub{R}{T}$, is better suited for practical applications. The final $\dSTO$ value is then determined by averaging $\dSTO$ over these 50 time steps to further mitigate numerical noise. 

In the simulated section of the magnetic pattern (with a length of $2m$ islets and buffers), we set $A(x) = A_0 = const$. This is necessary, because it allows the use of PBCs. Nevertheless, this introduces a small error to the calculation, because the SAW attenuation in this small section slightly changes $\pdiff{\vec{u}}{x}$ and thus $\boldsymbol{\varepsilon}$. However, this error is negligible when the decay of $A(x)$ within one wavelength of the SAW is small, as is the case there. The transmission efficiency itself is independent of the choice of $A$, as only the ratio between $\ssub{R}{T}$ and $\Energy{Ph}$ is relevant for its calculation (Eq. \ref{eq:dSijCalcTheory}) and both are proportional to $A^2$ in the linear regime of the LLG.

At $\ssub{f}{SAW} = 4.546\,\si{GHz} $, where the SAW wavelength is an integer multiple of the periodicity of the islets ($\lambda = \frac{3068.2\,\si{m/s}}{4.546\,\si{GHz}} = 3\cdot225\,\si{nm}$), conditions from Eqs. (\ref{eq:ValidationAbortRel}) or (\ref{eq:ValidationAbortAbs}) were not met within 10ns of simulation time for any device or configuration. Similarly, for $\ssub{f}{SAW} = 3.409\,\si{GHz} $, where $\lambda = \frac{3068.2\,\si{m/s}}{3.409\,\si{GHz}} = 4\cdot225\,\si{nm}$, the A-state of the 1D device did not converge (the other three simulations at that frequency did, however). For all five of these simulations, the instantaneous $\dSTO$ values oscillated slightly ($<\!0.4\,\si{dB}$ at \SI{4.546}{GHz} and $\approx\!1.6\,\si{dB}$ at \SI{3.409}{GHz}) around a constant value, where that value would be consistent with those of the surrounding frequencies. These simulations were nevertheless not included in the results. We attribute this to higher order harmonics also being excited by the SAW, which only get damped when their periodicity does not match that of the islets.

\subsection*{Interpretation}

\begin{figure}
    \centering
        \includegraphics[width=0.47\textwidth]{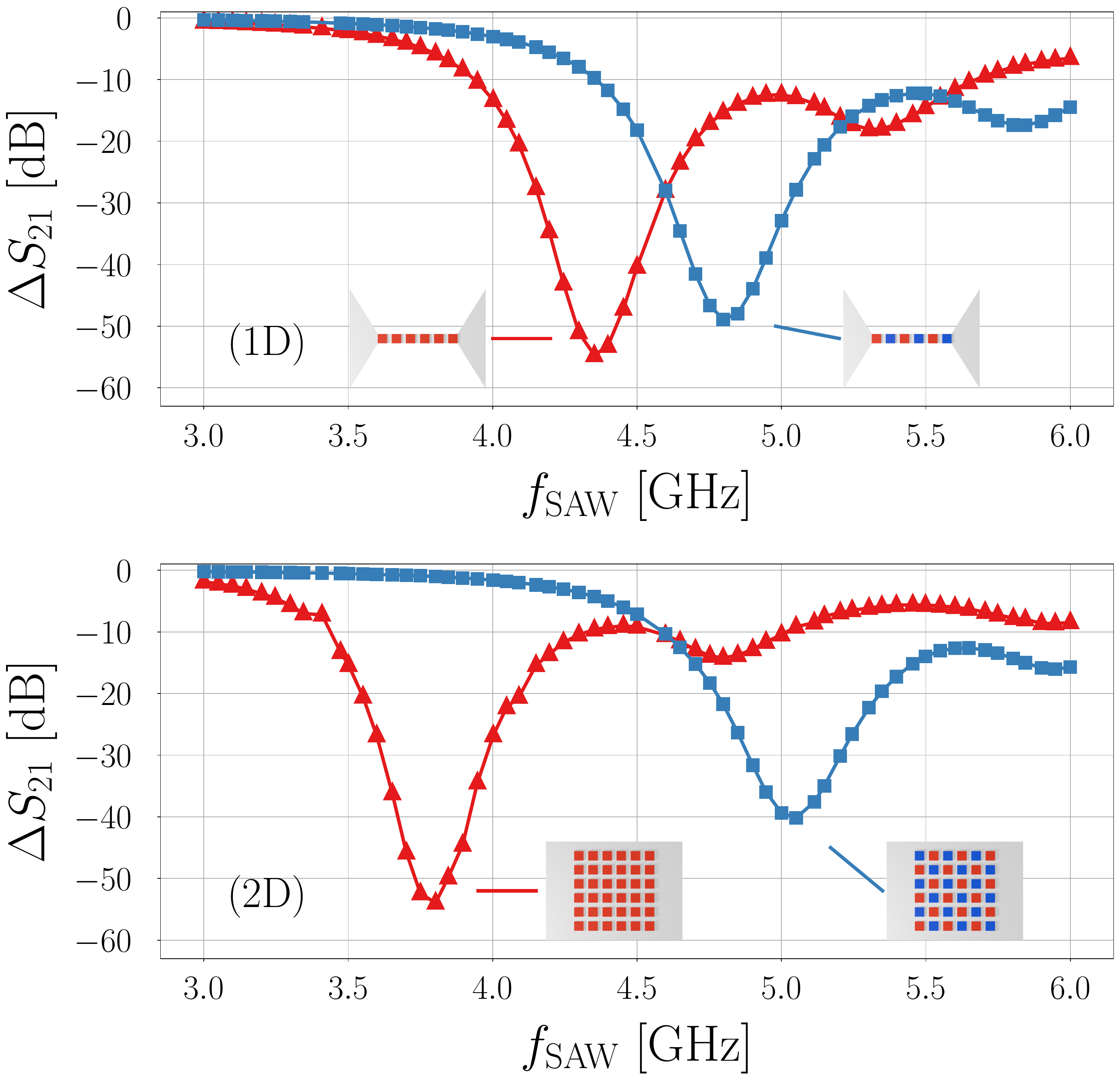}
        \caption{Simulation results for the transmission losses due to the magnon-phonon interaction, $\dSTO$, after \SI{1}{mm} travel distance in the 1D (top) and 2D (bottom) devices at different frequencies of the SAW in different magnetic states: parallel (red triangles) and antiparallel (blue squares). Insets show illustrations of the different islet arrangements. Simulations were performed under the influence of a \SI{50}{mT} bias field along the x direction.}
        \label{fig:Results}
\end{figure}

Figure \ref{fig:Results} shows the transmission losses $\dSTO$ of the SAW resulting from magnon-phonon interaction in both 1D and 2D devices, under parallel and antiparallel configurations. A notable shift in the location of the transmission dips is observed in the 1D and 2D devices (\SI{0.45}{GHz} and \SI{1.25}{GHz}, respectively), which largely correlates with the SW resonance modes presented in Fig. \ref{fig:Dispersion}, albeit at slightly higher frequencies than predicted. This deviation is attributed to both the broad response observed from the sinc pulse excitation as well as the islets not being homogeneously magnetized. The smaller transmission dips appearing at higher frequencies (e.g. \SI{4.80}{GHz} and \SI{5.95}{GHz} for the 2D P-state) are also in agreement with the detected SW modes between \SI{3}{GHz} and \SI{6}{GHz} in Fig. 3. Importantly, significant differences in SAW attenuation between the states are observed, which is optimized for our 2D device at \SI{3.80}{GHz}, with $\dSTO = -54.0\,\si{dB/mm}$ in the P-state and only \SI{-2.0}{dB/mm} in the A-state.

While the presented results are purely theoretical, all material parameters and geometries were chosen such that an experimental implementation should be feasible (see sec. Experimental Feasibility). Furthermore, we hypothesize the existence of configurations that allow for frequency absorption at points between simple parallel alignment or perfect checkerboard arrangements. If achievable, integrating this functionality with chirped IDTs, which are capable of exciting a broad range of frequencies rather than a single one \cite{Weiß2018Chirped}, would facilitate the development of a millimeter-sized RF notch filter with arbitrary frequency selectivity.

\section*{\label{sec: Discussion} Discussion}

In this study, we have demonstrated the theoretical viability of a magneto-phononic device concept capable of switching SAW attenuation at a target frequency solely depending on its internal magnetic configuration. The shift in the SW dispersion relation resulting from the stray-field interactions of the parallel vs antiparallel aligned magnetic Co/Ni islets making up the device was shown to be significant enough to alter the SAW attenuation by as much as \SI{52.0}{dB/mm} between them.

The algorithm developed to carry out these simulations was successfully validated on the results presented in ref. \cite{Kuess2021Experiment}, showing that the assumptions made in its derivation are reasonable. Furthermore, due to the reduction in computational complexity by this approach, it is possible to sweep through a very large set of simulation parameters in a reasonable amount of time while maintaining great accuracy.

\section*{Methods}

\subsection*{\label{sec: Uni-Dir Derivation} Uni-Directional Model: Derivation}

\begin{figure}
    \centering
        \includegraphics[width=0.47\textwidth]{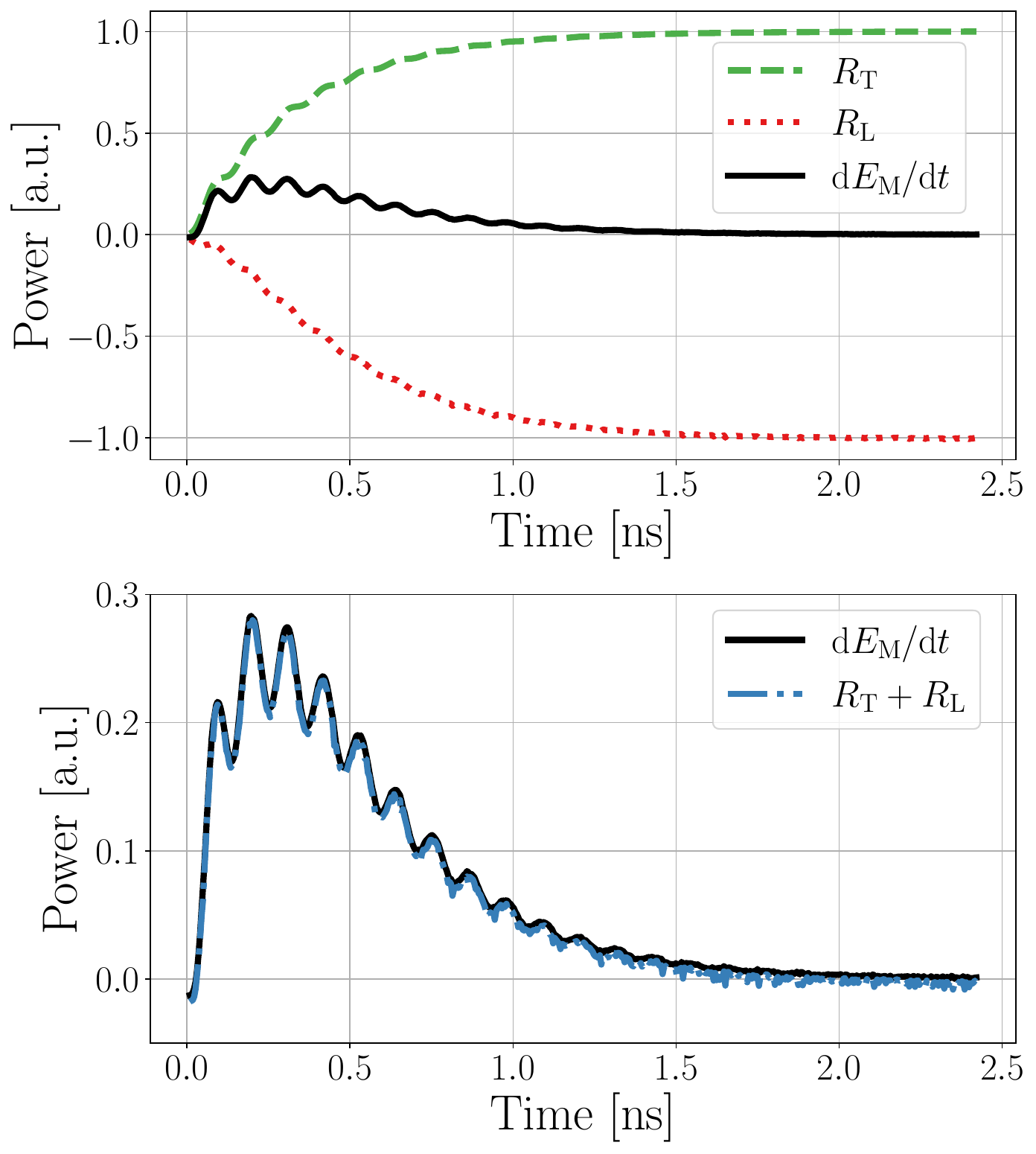}
    \caption{Example simulation of the magneto-phononic interaction within a thin magnetic film under influence of an external field. The system was in energetic equilibrium when at $t=0$ a continuous SAW signal was launched. The top graph shows the dissipative losses of the magnetic system (Eq. \ref{eq:RLDefinition}) in red, the energy transferred from the phonon to the magnon (Eq. \ref{eq:RTDefinition}) in green and the total change in energy in black $\brak{\text{approximated via}\, \Ediff{M}(t) \approx \frac{\ssub{E}{M}(t + \Delta t) - \ssub{E}{M}(t - \Delta t)}{2\Delta t}}$. At $t = 2.43\,\si{ns}$, the convergence criterion was reached. The bottom graph shows the validity of Eq. (\ref{eq:MSplit}) with $\ssub{R}{T} + \ssub{R}{L}$ (blue) indeed summing to $\Ediff{M}$ (black).} \label{fig:ExampleSim}
\end{figure}

The principal assumptions of the model are:

\begin{itemize}
    \item The SAW only loses energy through a reduction in its amplitude $A$. While it is known that the SAW velocity can change due to the magneto-phononic interaction, the relative change is typically limited to $<\!1\%$ \cite{Thevenard2014, Rovillain2022, Vythelingum2025}. As we want to calculate the magnetic transmission losses $\dSij$, where relative changes can reach $\approx\!100\%$, we neglect these small changes in $c$. In systems where a significant change of $c$ is expected, this would need to be incorporated in the model.
    \item Similarly, magneto-rotational coupling \cite{Xu2020, Rovillain2022, Vythelingum2025} is also not incorporated into our model, as its contributions are small when the saturation magnetostriction $\lambda_s$ is large (see Supplementary Information for details). 
    \item Measurements are only taken, once the system is in a state of constant energy flow between the mechanical and magnetic systems ($\diff{\ssub{R}{T}}{t} = 0$). This can only occur for a sustained SAW signal of at least a couple periods (see Fig. \ref{fig:ExampleSim}).
\end{itemize}

We start by examining the total change in energy of the magnetic system $\Ediff{M}$:

\begin{align} 
    \frac{\text{d}E_{\text{M}}}{\text{d}t} &= \IntO{ \frac{\text{d}}{\text{d} t} U_{\text{M}}\bbrak{t, \vec{m}(t)}} \\
    &= \underbrace{\IntO{\frac{\delta U_{\text{M}}\bbrak{t, \vec{m}(t)}}{\delta \textit{\textbf{m}}} \frac{\text{d}\textit{\textbf{m}}}{\text{d}t}}}_{\equiv \ssub{R}{L}} + \underbrace{\IntO{\frac{\partial U_{\text{M}}\bbrak{t, \vec{m}(t)}}{\partial t}}}_{\equiv \ssub{R}{T}}. \label{eq:MSplit}
\end{align}

\noindent Here, $\ssub{U}{M}$ denotes the magnetic energy density and $\ssub{R}{L}$ are the dissipative losses due to Gilbert damping with the well known result \cite{Abert2019MicroMagReview}

\begin{align}
    \ssub{R}{L} &=- \frac{\alpha\gamma}{1 + \alpha^{2}}\mu_{0}M_{\text{s}}\IntO{ (\textit{\textbf{m}} \times \textit{\textbf{H}}_{\text{Eff}})^{2} }. \label{eq:RLDefinition}
\end{align}

For $\ssub{R}{T}$, we make use of the fact that, for constant $\Hext$, the magneto-elastic energy term (Eq. \ref{eq:DefinitionMagElEnergy}) is the only one that explicitly depends on time in $\ssub{E}{M}$. When the simulation geometry is an integer multiple of the wavelength, $\IntO{\pdiff{}{t} \brak{\boldsymbol{\varepsilon} : C : \boldsymbol{\varepsilon}}} = 0$. Given that $\ssub{\boldsymbol{\varepsilon}}{m}$ has no explicit time dependence, $\pdiff{\ssub{\boldsymbol{\varepsilon}}{m}}{t} = 0$. Thus, for a symmetric $C$, $\ssub{R}{T}$ simplifies to

\begin{align}
    \ssub{R}{T} &= - \IntO{\brak{\pdiff{\boldsymbol{\varepsilon}}{t}} : C : \boldsymbol{\varepsilon}_{\text{m}}}. \label{eq:RTDefinition} 
\end{align}

Using the law of energy conservation, we know that all energy gained by the magnetic system has to be compensated by a loss in the energy of the phonon, $\Energy{Ph}$:

\begin{align} \label{eq:EnergyPhononIsGain}
    \Ediff{Ph} = -\ssub{R}{T} = \IntO{\brak{\pdiff{}{t} \boldsymbol{\varepsilon}} : C : \boldsymbol{\varepsilon}_{\text{m}}}.
\end{align}

If the magneto-phononic interaction is weak enough, that the LLG stays in a linear regime, it is known, that $\Energy{Ph}$ follows an exponential decay \cite{Dreher2012Theory, Kuess2021Experiment} (in the Supplementary Information, a possible extension of this approach to the non-linear regime is discussed):

\begin{align}
    \ssub{E}{Ph}(t) &= \ssub{E}{Ph}(0) e^{\beta t}. \label{eq:ExponentialAnsatz}
\end{align}

Using Eq. \ref{eq:EnergyPhononIsGain}, we can express the decay rate $\beta$ as

\begin{align}
    \beta = \beta \frac{\ssub{E}{Ph}(0) e^{\beta t}}{\ssub{E}{Ph}(0) e^{\beta t}} &= \frac{1}{\ssub{E}{Ph}(t)} \Ediff{Ph}(t) = \frac{-\ssub{R}{T}(t)}{\ssub{E}{Ph}(t)}
    \label{eq:BetaCalc} 
\end{align}

\noindent where we can see, that it is sufficient to know $\ssub{R}{T}$ and $\Energy{Ph}$ for any single point in time, in order to obtain $\beta$ for the full interaction. $\ssub{R}{T}$ can be calculated from Eq. \ref{eq:RTDefinition}, where we simulate a segment of the magnetic material until $\ssub{R}{T}$ becomes constant (see fig \ref{fig:ExampleSim}). $\Energy{Ph}$ can be obtained in a pre-simulation step by summing the potential energy $\frac{1}{2} \boldsymbol{\varepsilon} : C : \boldsymbol{\varepsilon}$ and kinetic energy $\frac{\rho \vec{v}^{2}}{2}$ of the SAW over the full volume $V$, including possible non-magnetic layers \cite{Landau1960TOE}:

\begin{align}
    \ssub{E}{Ph} = \frac{1}{2} \int_V \boldsymbol{\varepsilon} : C : \boldsymbol{\varepsilon} + \rho \vec{v}^{2} \dxx. \label{eq:TotalSAWEnergy}
\end{align}

\noindent Given that both $\boldsymbol{\varepsilon}$ and $\vec{v}$ are derived from \vec{u}, they are proportional to $A$ and therefore: $\ssub{E}{Ph} \propto A^2$. Using $P \propto A^2$ \cite{Robbins1977SAWPower}, where $P$ is the power of the SAW, we find:

\begin{align}
    P \propto \ssub{E}{Ph}.
\end{align}

Finally, by viewing the phonon as a quasi-particle with a position $x(t) = c t$, where $c = \ssub{v}{t}\xi$ is its velocity, we can give its energy in terms of its position instead of time:

\begin{align}
    \ssub{E}{Ph}(t) = \ssub{E}{Ph}\brak{\frac{x(t)}{c}}. \label{eq:SpaceToTime}
\end{align}

Combining Eq. \ref{eq:EnergyPhononIsGain} to \ref{eq:SpaceToTime}, we obtain the magnetic transmission losses $\dSij$ by setting the position of the phonon $x(t)$ to the length of the magnetic material $l$:

\begin{align}
    \Delta S_{ij}(l) &= 10 \log{\frac{\ssub{P}{out}}{\ssub{P}{in}}} \\[0.5em]
    &= 10 \log{\frac{\ssub{E}{Ph}(l/c)}{\ssub{E}{Ph}(0)}} \\[0.5em]
    &= 10 \log{\exp{\frac{l}{c} \beta}}\\[0.5em]
    &= \frac{10}{\ln{10}} \frac{l}{c} \beta \\
    &= \frac{10}{\ln{10}} \frac{l}{c} \frac{-\ssub{R}{T}}{\ssub{E}{Ph}} \\
    &= \frac{10}{\ln{10}} \frac{l}{c} \frac{\IntO{\brak{\pdiff{}{t} \boldsymbol{\varepsilon}} : C : \boldsymbol{\varepsilon}_{\text{m}}}}{\frac{1}{2} \int_V \boldsymbol{\varepsilon} : C : \boldsymbol{\varepsilon} + \rho \vec{v}^{2} \dxx}. \label{eq:dSijCalc}
\end{align}

\subsection*{\label{sec: Uni-Dir Validation} Uni-Directional Model: Validation}

\begin{figure} [H]
    \centering
        \includegraphics[width=0.47\textwidth]{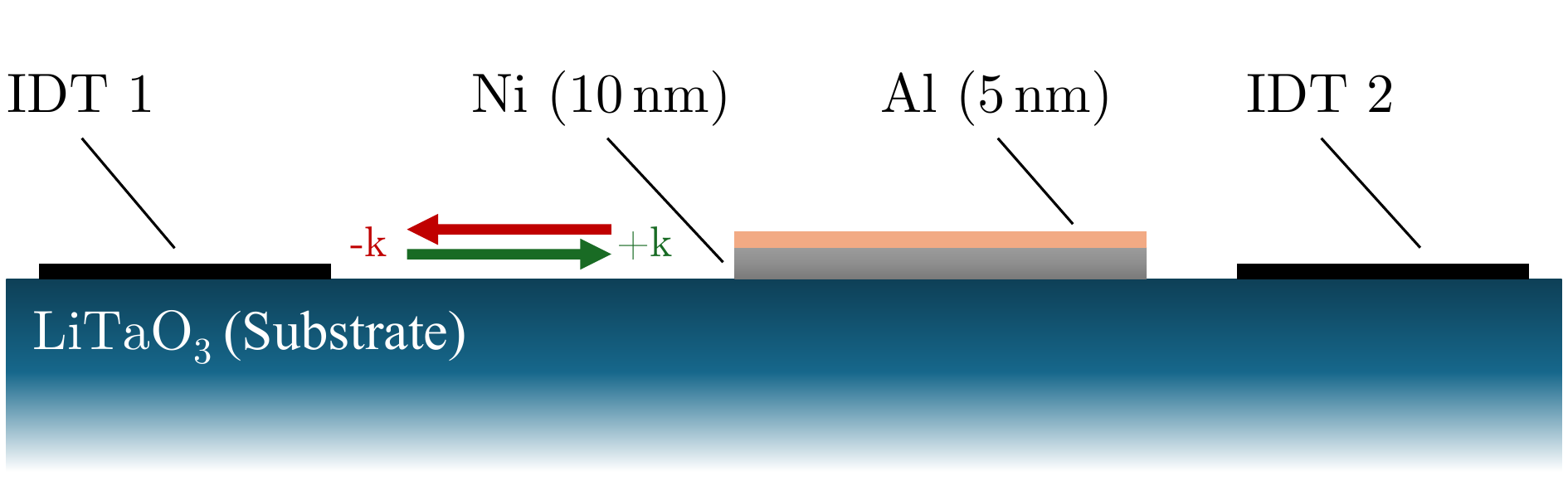}
        \caption{Illustrated cross-section of the device employed in ref. \cite{Kuess2021Experiment}. On top of a 36°-rotated Y-cut X-propagation LiTaO$_\text{3}$ substrate, two inter digital transducers (IDTs) are placed $1600\,\text{µm}$ apart. Between them, a $1000\,\text{µm}$ long and \SI{10}{nm} thick Ni film is applied and capped with a \SI{5}{nm} Al layer. $+\vec{k}$ and $-\vec{k}$ indicate the direction of SAW travel.}
        \label{fig:KuessExpSetup}
\end{figure}

To validate the model, we utilize material parameters from ref. \cite{Kuess2021Experiment}, derived from both experimental measurements and fitting procedures. These parameters characterize a layered system comprising a \SI{10}{nm} Ni thin film deposited on a LiTaO$_3$ substrate, capped with a \SI{5}{nm} Al layer (see Fig. \ref{fig:KuessExpSetup}). We then compare the simulated transmission losses of a Rayleigh SAW with corresponding experimental results. The validation procedure is largely identical to the one introduced in sec. SAW parameterization and Magneto-Phononic Interaction, only now the frequency of the SAW gets fixed at $\ssub{f}{SAW}=4.47\,\si{GHz}$, while the external field is varied. The following energy terms were considered: (i) demagnetization energy, (ii) exchange energy with stiffness $\ssub{A}{ex}$, (iii) in-plane uniaxial magnetic anisotropy $\ssub{K}{u}^{\text{IP}}$ in easy axis direction $\ssub{\phi}{u}^{\text{IP}}$, (iv) OOP surface magnetic anisotropy counteracting the shape anisotropy $\ssub{K}{u}^{\text{OOP}}$, (v) magneto-elastic energy with saturation magnetostriction $\ssub{\lambda}{s}$, and (vi) Zeeman energy of an external field of strength $\Hext$ and angle $\phiH$. $\ssub{\phi}{u}^{\text{IP}}$ and $\phiH$ are given in-plane with the x-axis. Parameters for the Ni film are taken from ref. \cite{Kuess2021Experiment} and can be seen in Table \ref{tab:MagParasKuess}. 

\begin{table}[H]
\centering
    \caption{Magnetic parameters of the \SI{10}{nm} thick Ni film. \cite{Kuess2021Experiment}}
    \begin{tabular}{lcccccc}
        \toprule
        $\alpha$ & $\ssub{M}{s}$ & $\ssub{A}{ex}$ & $\ssub{K}{u}^{\text{IP}}$ & $\ssub{\phi}{u}^{\text{IP}}$ & $\ssub{K}{u}^{\text{OOP}}$ & $\ssub{\lambda}{s}$ \\
        $[1]$ & [kA/m] & [pJ/m] & [kJ/m$^3$] & $[^\circ]$ & [kJ/m$^3$] & $[10^{-6}]$ \\
        \midrule
         0.069 & 408.0 & 7.7 & 0.28 & 83.6 & 23.8 & -14.22  \\
        \bottomrule
    \end{tabular}
    \label{tab:MagParasKuess}
\end{table}

\noindent Here, $\ssub{K}{u}^{\text{OOP}}$ was fitted, such that the peak absorption for $\phiH = 45^\circ$ occurs at the same $\mu_0\Hext$ as the experiment (\SI{-46}{mT}). $\ssub{\lambda}{s}$ was then determined by matching simulation results for $\dSTO\brak{\mu_0\Hext=-46\,\si{mT},\,\phiH = 45^\circ}$ with the experimental value. This was done, because $\ssub{\lambda}{s}$ was not a fit parameter in ref. \cite{Kuess2021Experiment} and $\ssub{K}{u}^{\text{OOP}}$ was fitted in conjunction with an analytical approximation of the stray field in ref. \cite{Kuess2021Experiment}, whereas the micromagnetic energy term given in Eq. (\ref{eq:DefEnergy}) is used for the simulations here. Note that the fitted $\ssub{K}{u}^{\text{OOP}}$ of \SI{23.8}{kJ/m^3} shows good agreement with the measured $\ssub{K}{u}^{\text{OOP}}$ of \SI{32.7}{kJ/m^3} obtained from broadband ferromagnetic resonance measurements \cite{Kuess2021Experiment}.

During validation, $\mu_0 \Hext$ was varied from \SI{-72}{mT} to \SI{72}{mT} in increments of \SI{2}{mT} and $\phiH$ from $-90^\circ$ to $90^\circ$ in increments of $4.5^\circ$ for both directions of travel of the SAW, for a total of 5986 individual simulations. For each, the system was initialized with $\vec{m}(\vec{x}) = (\text{sgn}(\Hext) \cos{\phiH}, \text{sgn}(\Hext) \sin{\phiH}, 0)$ and relaxed before the SAW was activated.

\begin{figure}
    \centering
        \includegraphics[width=0.47\textwidth]{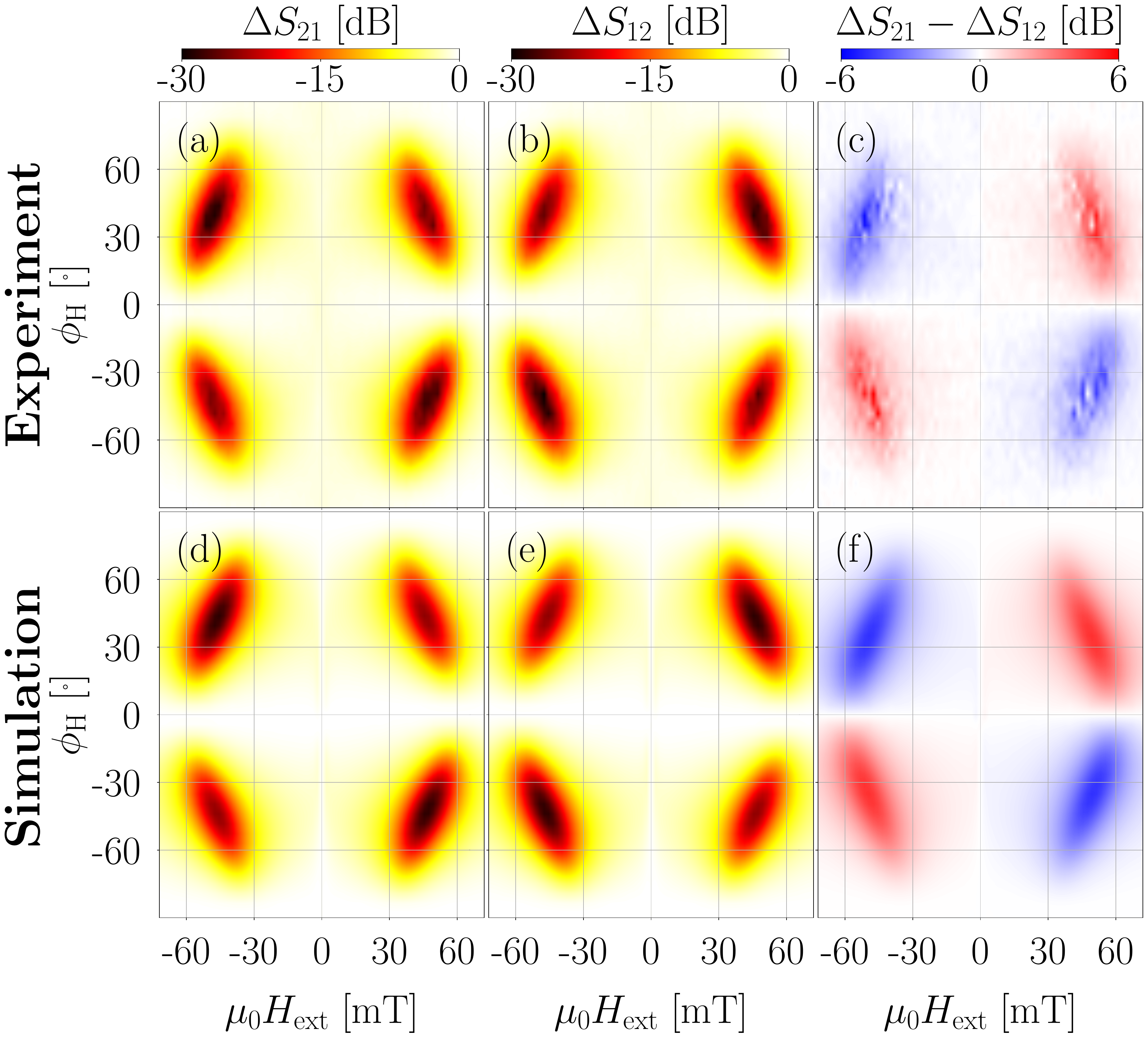}
    \caption{(a) to (c): Experimental results of ref. \cite{Kuess2021Experiment}. (d) to (f): Results of micromagnetic simulations. (a), (b), (d), and (e) show the transmission losses due to the magneto-phononic interaction while (c) and (f) give the non-reciprocity between the travel directions.}
    \label{fig:KuessComparison}
\end{figure}

The results of this validation run can be seen in Fig. \ref{fig:KuessComparison}. The micromagnetic simulation and the experimental results show excellent agreement not only qualitatively but also quantitatively for both directions of SAW travel. The expected four-fold symmetry is clearly present, as well as the non-reciprocity between the two directions of travel. The validation took a total of $\approx\!100\,\si{hours}$ on an NVIDIA A100 (80GB) GPU, for an average of only $\approx\!1\,\si{minute}$ per simulation.

\subsection*{\label{sec: ExpFeas}Experimental Feasibility}

To transition the theoretical potential of this device toward practical application, we outline potential solutions for its experimental execution here.

While there are multiple possible venues to fabricate the islets, FIB seems particularly promising, as this approach has already been heavily investigated in regards of patterning a continuous thin film \cite{Rettner2001}.  With this method, single-domain islands with a side-length of \SI{70}{\nano \meter} and a distance of \SI{25}{\nano \meter} were achieved \cite{Rettner2001}. Another possibility would be to first deposit a continuous thin film, then use e-beam-lithography and reactive ion etching to pattern the film \cite{Krone2011, Grobis2011BPM}. Here, islets had a diameter as small as \SI{40}{\nano \meter} with a \SI{60}{\nano \meter} gap. \cite{Krone2011}.

To perform a sensitivity analysis of the proposed device, we varied the following material parameters from half to double their assumed magnitude (see Table \ref{tab:DeviceParameters}) in the 2D A-state, revealing the following relations: $\dSij \propto \lambda_s^2$, $\dSij \propto E^{2}$ and $\dSij \propto \frac{1}{(1+\nu)^2}$. Variations upon $\alpha$ led to changes in the line width of the absorption while simultaneously scaling the peaks according to $\dSij \propto \alpha^{-1}$. Finally, $\ssub{A}{ex}$ and $\Ku$ determine the frequency of most efficient coupling for both P- and A-states as they shift the SW dispersion relation, while $\Ms$ influences the magnitude of the shift between these coupling frequencies.

To program the device in experiments, we suggest first saturating the islets along the z-direction, thereby achieving the P-state. Starting from this state, by applying a field in the opposite direction, we expect individual islets to start flipping due to small inhomogeneities of $\Ku$ present in a real sample. By increasing the field strength, more islands switch. Therefore, an arbitrary number of islands can be switched depending on the field strength. To achieve the A-state, every other island could be irradiated minimally in order to slightly reduce the anisotropy of these islets. We anticipate this would then lead to them flipping before islets which were not irradiated, leading to an almost perfect checkerboard pattern.

Spin-torque methods, such as those utilized in Spin-Transfer Torque and Spin-Orbit Torque MRAM technologies \cite{Guo2021STTSOTRoadmap} represent an alternative approach to programming the islets. Implementation of these techniques would likely necessitate more complex geometric structures and intricate layer stacks than those examined in the current study and are, therefore, excluded from our present scope. Nevertheless, the successful integration of spin-torque switching could facilitate dynamic, individual addressing of the magnetic islets, thereby enabling the realization of more complex magnetization patterns beyond the simple P- and A-states considered here - possibly enabling additional functionality.

\section*{Data Availability}

The data that support the plots presented in this paper are available from the corresponding author upon reasonable request.

\section*{Code Availability}

The simulation files used to generate and analyze the data are available from the corresponding author upon reasonable request.

\bibliography{main}

\section*{Acknowledgement}

This work was funded in whole or in part by the Austrian Science Fund (FWF) [DOI: 10.55776/I6068 \& 10.55776/P34671] and by the Deutsche Forschungsgemeinschaft (DFG, German Research Foundation) project No. 504150161. We acknowledge the financial support by the Vienna Doctoral School
in Physics (VDSP). For open access purposes, the author has applied a CC BY public copyright license to any author-accepted manuscript version arising from this submission.

\section*{Contributions}

D.S., H.J.K., M.A. and C.A. conceived the islet design. M.K.S. and C.A. formulated the uni-directional model. M.K.S performed and analyzed all micromagnetic simulations. E.D.S.N. and M.W. carried out all finite element simulations. P.F. implemented the magneto-elastic solver in magnum.np and analyzed the finite element simulations. M.K. performed the experiment used for the validation of the uni-directional model. M.K., S.G., H.J.K. and M.A. guided the design of the device with their experimental expertise. D.S., H.J.K., M.A. and C.A. jointly supervised the work. M.K.S. wrote the original draft with help from all other authors.

\end{document}